\documentclass[aps,pra,twocolumn,10pt]{revtex4-1}
\usepackage[utf8]{inputenc}
\usepackage{amsmath,amssymb,graphicx,bm,microtype}
\usepackage[cal=pxtx]{mathalfa}
\usepackage[colorlinks=true,citecolor=blue,urlcolor=blue]{hyperref}

\begin{document}
\title{Singlet-triplet transition in double quantum dots in two-dimensional topological insulators}

\author{Vladimir~A. Sablikov and Aleksei~A. Sukhanov}
\affiliation{Kotel’nikov Institute of Radio Engineering and Electronics, Russian Academy of Sciences,
Fryazino, Moscow District, 141190, Russia}

\begin{abstract}
We study two-electron states confined in two coupled quantum dots formed by a short-range potential in a two-dimensional topological insulator. It is shown that there is a fairly wide range of the system parameters, where the ground state is a tripletlike state formed by a superposition of two spin-polarized states. Outside this range, the ground state is a singlet. A transition between the singlet and triplet states can be realized by changing the potential of the quantum dots. The effect is caused by a significant change in the energies of the Coulomb repulsion and the exchange interaction of electrons due to the presence of the pseudospin components of the wave function when the band spectrum is inverted.
\end{abstract}

\maketitle

\section{Introduction}
\label{Intro}
The problem of the ground state spin of a quantum system of interacting electrons has a long history and is of great importance nowadays because of the prospects of creating spin qubits as a compact platform for quantum computations~\cite{doi:10.1146/annurev-conmatphys-030212-184248,Awschalom1174}. The spin state is formed mainly due to the exchange correlations of the electrons and their interaction with confining potential. A definite answer to the question of the ground-state spin exists in the case of two electrons. There is a theorem that states that the ground state is a singlet if no spin- or velocity-dependent forces are present~\cite{mattis1965theory}. As an exact result, this theorem is important for understanding the quantum states of interacting fermions. But its proof is restricted by considering the wave functions in the form of Pauli second-rank spinors, which is generally not the case in modern materials, where the wave-function spinors contain also pseudospin components describing the orbital degrees of freedom. It is this problem that we address in the present paper.  

The above theorem is generalized to the many-particle system only in one dimension~\cite{PhysRev.125.164}. For few-electron and multielectron systems in two and three dimensions there is no such strict theorem; however there is the famous semiempirical Hund’s multiplicity rule established for multielectron atoms. It states that the lowest energy has the term with the highest spin, which is possible for a given electron configuration, and with the highest orbital moment at this spin~\cite{landau1977quantum}. The origination of Hund’s rule has been debated for a long time~\cite{boyd1984quantum,kutzelnigg1996hund}, but there is no rigorous proof of its validity conditions. In particular, it is not clear how many electrons should be in a confined system for its ground state to be a triplet.

The situation changes when a confined electron system is coupled to a many-electron system. Thus, the ground state of a quantum dot (QD) can be a triplet when it is coupled to electron reservoirs in the Kondo-effect regime at even filling~\cite{PhysRevB.67.113309}. A triplet ground state can be formed in a small QD coupled to another multielectron QD, which serves as an exchange mediator~\cite{PhysRevLett.114.226803,PhysRevLett.119.227701,PhysRevX.8.011045,PhysRevB.97.245301}.

In the presence of a magnetic field, the singlet-triplet (S-T) transition can occur in an isolated quantum dot with two electrons~\cite{PhysRevB.45.1951}. Great interest is paid in the literature to double-QD structures where in the presence of a magnetic field the S-T transition is controlled electrically by changing the voltage between the QDs~\cite{PhysRevA.57.120,RevModPhys.79.1217}.

In this paper we draw attention to the fact that above results were obtained for confined electron systems hosted in materials with the usual band spectrum. In topologically nontrivial materials, the situation changes greatly due to the presence of additional orbital degrees of freedom. As a consequence, the wave function is a higher-rank spinor that includes both spin and pseudospin components. Because of this, first, the exchange interaction changes essentially since not only the spins but also the pseudospins are rearranged when the particles are permuted. Second, due to the pseudospin components the spatial distribution of the electron density changes, and consequently, the electron interaction with the confining potential of the QDs also changes. Topology-dependent effects due to the pseudospin components arise already in the interaction of a single electron with a localized potential, as has been demonstrated in spectra of single-particle states bound to an impurity~\cite{1367-2630-13-10-103016,doi:10.1002/pssr.201409284,PhysRevB.91.075412,PhysRevB.92.085126,SUKHANOV20171}. In the case of two-particle states, \textit{a fortiori}, the appearance of nontrivial effects due to pseudospins can be expected.

This paper aims to elucidate the spin states of two interacting electrons confined in double QDs in topologically nontrivial materials. We study two-electron states confined in two coupled narrow quantum wells in a two-dimensional (2D) material with a two-band spectrum described by the  Bernevig-Hughes-Zhang (BHZ) model~\cite{BHZScience2006}. We have found very unusual properties of the spectrum and spin structure of states. The main feature is that in the topological phase the ground state can be either singlet or triplet depending on the system parameters. The transition between the singlet and triplet states can be realized by changing the potential of the wells.

\section{Double-quantum-well model}\label{Sec2}
Consider two coupled QDs formed by a double-well potential in a 2D material described by the BHZ model. The Hamiltonian is
\begin{equation}
H=H_{BHZ}+V_A(|\mathbf{r}-\mathbf{R}_A|) + V_B(|\mathbf{r}-\mathbf{R}_B|)\,,
\label{H}
\end{equation}
where $V_A$ and $V_B$ are potentials of the QDs and $R_A$ and $R_B$ stand for their positions. $H_{BHZ}$ is the standard BHZ Hamiltonian~\cite{BHZScience2006}, which we take in symmetric form with respect to the electron and hole bands. In this work we neglect the spatial inversion asymmetry, so that the one-particle Hamiltonian is block diagonal.  

Bound eigenstates of the Hamiltonian~(\ref{H}) can be found analytically in the case where the potentials $V_A$ and $V_B$ are short-range ones. We calculate the wave functions of the bound states using the Fourier transform. This approach was used in our recent works~\cite{doi:10.1002/pssr.201409284,PhysRevB.91.075412} for a single quantum well with a short-range potential. Here we generalize it for double wells. The application of this method allows us to describe the coupling of the wells nonperturbatively. Since the system is symmetric with respect to $S_z$, it is enough to consider one of the spin sectors where the wave function is a spinor $\Psi=(\psi_1, \psi_2)^T$ in the basis $(|e\rangle, |h\rangle)^T$, where $|e\rangle$ and $|h\rangle$ are the basis states of the electron and hole bands. The Schr\"odinger equation has the form
\begin{equation}
[\varepsilon-h(\hat{\mathbf{k}})]\Psi(\mathbf{r})=[V_A(|\mathbf{r}-\mathbf{R}_A|) + V_B(|\mathbf{r}-\mathbf{R}_B|)]\Psi(\mathbf{r}),
\label{Schrodinger_1}
\end{equation}
where dimensionless units are used: the energy and potentials are normalized to the mass term in the BHZ model $|M|$; the distance is normalized to $\sqrt{|M/B|}$, with $B$ being the band dispersion parameter. The operator $h(\hat{\mathbf{k}})$ is
\begin{equation} 
h(\mathbf{k})=
\begin{pmatrix}
 \mu+k^2 & a(k_x+ik_y)\\
 a(k_x-ik_y) & -\mu-k^2
\end{pmatrix}\,,
\end{equation}
with $a=A/\sqrt{|MB|}$ being the band hybridization parameter. The parameter $\mu=M/|M|$ is introduced to separate the topological ($\mu=-1$) and trivial ($\mu=1$) phases. The energy $\varepsilon$ is measured from the middle of the band gap.

Carrying out the Fourier transform of the right-hand side of Eq.~(\ref{Schrodinger_1}) we suppose that the potentials $V_{A,B}$ are localized at a distance shorter than the localization length of the wave function. Therefore  
\begin{multline}
\int d^2r V_{A,B}(|\mathbf{r}-\mathbf{R}_{A,B}|)\Psi(\mathbf{r})e^{-i\mathbf{k r}} \\ \approx \Psi_{A,B}e^{-i\mathbf{k R}_{A,B}} \widetilde{V}_{A,B}(k)\,,
\label{short_approx}
\end{multline}
where $\Psi_{A,B}=\Psi(\mathbf{R}_{A,B})$ and $\widetilde{V}_{A,B}(k)$ is the Fourier transform of $V_{A,B}(r)$.

Omitting the details, we arrive at the following results. The wave-function spinor has the form
\begin{equation}
\Psi(\mathbf{r})= \mathbf{G}_A(\varepsilon,\mathbf{r}-\mathbf{R}_A)\Psi_A + \mathbf{G}_B(\varepsilon,\mathbf{r}-\mathbf{R}_B)\Psi_B\,,
\label{single_WF}
\end{equation}
where $\mathbf{G}_A(\mathbf{r}-\mathbf{R}_A)$ and $\mathbf{G}_B(\mathbf{r}-\mathbf{R}_B)$ are the following matrices:
\begin{multline}
\mathbf{G}_{A,B}(\varepsilon,\mathbf{r}-\mathbf{R}_{A,B})\\= \int \!\frac{d^2k}{4\pi^2}\frac{\widetilde{V}_{A,B}(k)}{\Delta(\varepsilon,k)}
\begin{pmatrix}
 \varepsilon\!+\!\mu\!+\!k^2 & a(k_x\!+\!ik_y)\\
 a(k_x\!-\!ik_y) &  \varepsilon\!-\!\mu\!-\!k^2 
\end{pmatrix}
e^{i\mathbf{k}(\mathbf{r}-\mathbf{R}_{A,B})}\,,
\label{G_matrices}
\end{multline}
and $\Delta(\varepsilon,k)=\varepsilon^2-(\mu+k^2)^2-a^2k^2$.

The spinors $\Psi_A$ and $\Psi_B$ are determined by the equations
\begin{equation}
 \left\{ 
 \begin{array}{cl}
 [1-\mathbf{G}_A(\varepsilon,0)]\Psi_A - \mathbf{G}_B(\varepsilon,-\mathbf{R})\Psi_B=&0\\
-\mathbf{G}_A(\varepsilon,\mathbf{R})\Psi_A + [1-\mathbf{G}_B(\varepsilon,0)]\Psi_B=&0\,,
\end{array}\right.
\label{psi_AB}
\end{equation}
where $\mathbf{R}=\mathbf{R}_B-\mathbf{R}_A$.

Having been written for the components of spinors $\Psi_A$ and $\Psi_A$ explicitly, Eq.~(\ref{psi_AB}) represents the system of four linear equations that fully determines the spinor components at points $\mathbf{R}_A$ and $\mathbf{R}_B$ and the spectrum of the bound states. The equations are easily solved in a standard way. As a result, all quantities (the spectrum and spinor components) are expressed in terms of the components of the matrices $\mathbf{G}_{A,B}$, which are straightforwardly calculated by numerical integration.

In this way we have studied the spectrum of two identical QDs in both the topological and trivial phases for a variety of distances $d$ between the QDs and their potentials. The potential shape of the quantum wells is approximated by the Gaussian function $V(r)=(v\Lambda^2/\pi) \exp(-\Lambda^2r^2)$ or by the $\delta$ function $V(r)=(v/\pi)\delta(r^2)$. In the latter case, some integrals, which are logarithmically divergent, should be regularized by cutting off at $\Lambda$ when integrating with respect to $k$. In both cases the value $\Lambda$ should be large enough for Eq.~(\ref{short_approx}) to be satisfied. Numerical calculation shows that Eq.~(\ref{short_approx}) is well satisfied for the Gaussian potential if $\Lambda>2$.

\begin{figure}
\centerline{\includegraphics[width=.9\linewidth]{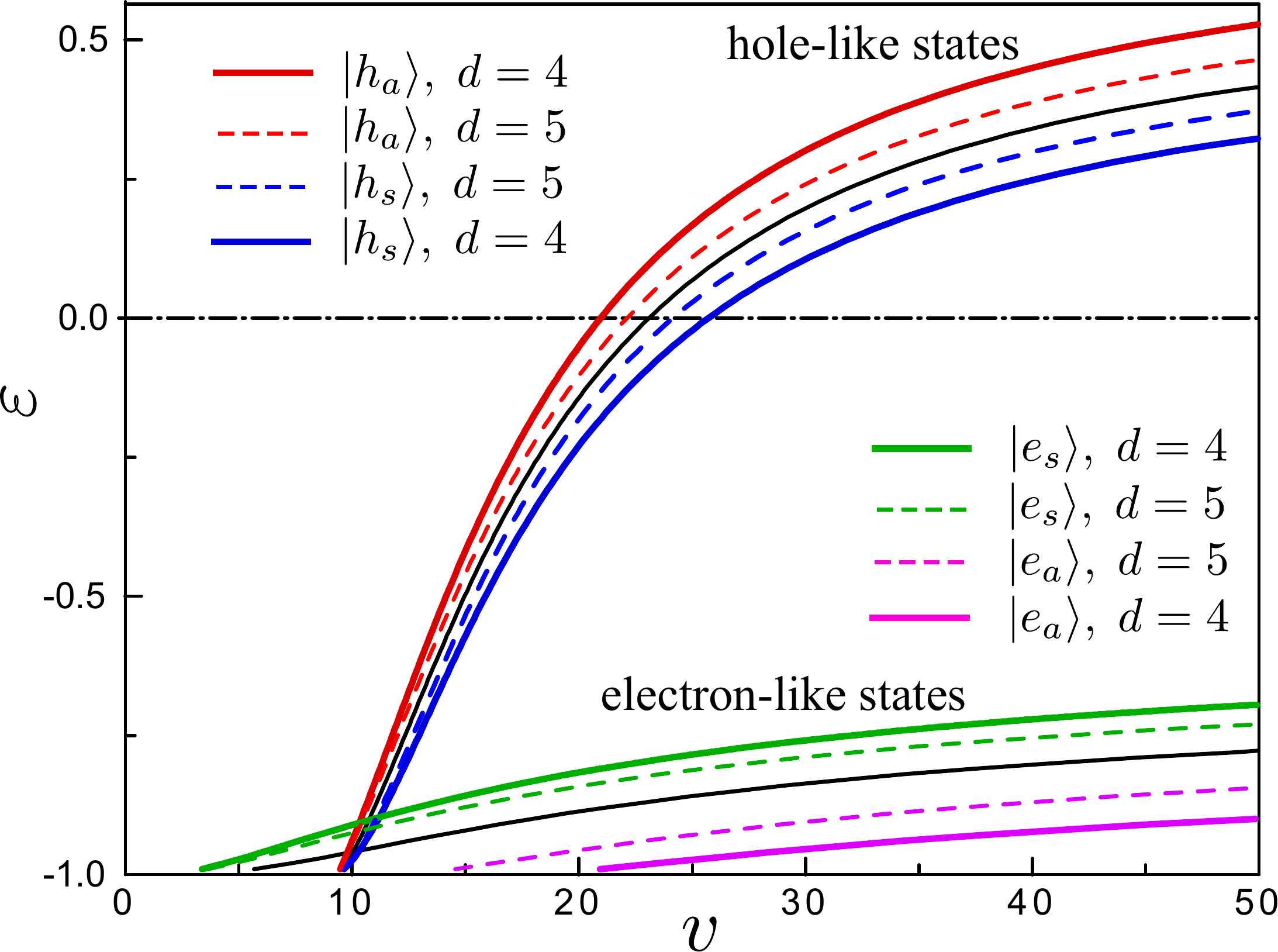}}
\caption{(Color online) Bound-state energy as a function of the quantum well potential for two interwell distances ($d=4$ and $d=5$) in the topological phase. Thin black lines depict the single-well energy spectrum. The parameters used in the calculations are $a=2$, $\Lambda=3$.}
\label{fig1}
\end{figure}

\begin{figure}
\centerline{\includegraphics[width=.8\linewidth]{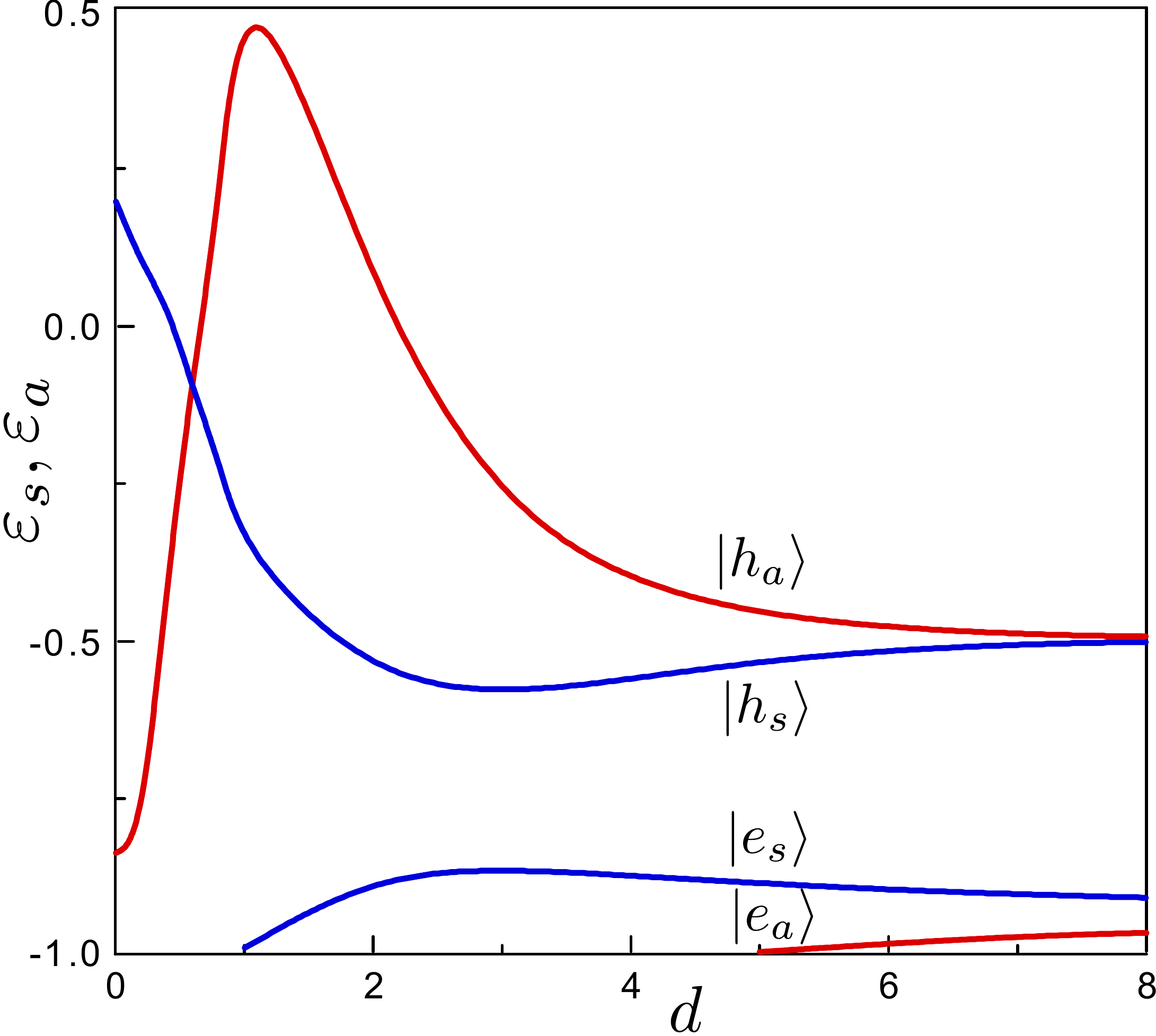}}
\caption{(Color online) Bound-state energy of the symmetric $\varepsilon_s$ and antisymmetric $\varepsilon_a$ states as a function of the interwell distance in the topological phase. Red and blue lines show, respectively, the symmetric and asymmetric electron- and holelike states. Numerical parameters are chosen as $v=15$, $\Lambda=3$, and $a=2$.}
\label{fig2}
\end{figure}

The main features of the spectrum are demonstrated in Figs.~\ref{fig1} and \ref{fig2} for the topological phase. First, note that the spectrum of a single quantum well formed by a short-range potential contains two states in the topological phase, in contrast to the trivial phase where there is only one state~\cite{doi:10.1002/pssr.201409284,PhysRevB.91.075412}. The states differ in their pseudospin structure. In one of the states (we call it the electronlike one), in the center of the well, $\psi_1\ne 0$ and $\psi_2=0$. In the other, the holelike state, on the contrary, $\psi_1= 0$ and $\psi_2\ne 0$ in the center.

The spectrum of the double wells is formed from this single-well spectrum by splitting it into two states. One of them is symmetric and the other is antisymmetric with respect to the inversion of the axis passing through the points of the well positions. Thus, there are two holelike and two electronlike states for each spin. We denote them as $|h_s\rangle$, $|h_a\rangle$, $|e_s\rangle$, and $|e_a\rangle$. The dependence of the bound-state energies on the well potential $v$ for the split states (Fig.~\ref{fig1}) is qualitatively similar to that of the single well. Figure~\ref{fig1} shows the spectra only for positive potential. For negative potential $v<0$, the bound states also exist, and their spectrum is obtained from that for $v>0$ by the symmetry relation: $\varepsilon(-v)=-\varepsilon(v)$.

The splitting between the symmetric and antisymmetric states strongly increases with decreasing interwell distance when $d$ is large. However, as $d$ is comparable with the length of the wave-function penetration within the interwell region, the bound-state energies vary with $d$ nonmonotonically, as shown in Fig.~\ref{fig2}. 

It is interesting that the energy difference between the antisymmetric and symmetric states can be either positive or negative depending on the type of states and the sign of the well potential. If $v>0$, the energy difference is positive for the holelike states and negative for the electronlike states. At $v<0$ the situation changes to the opposite. 

The wave functions are calculated using Eqs.~(\ref{single_WF}), (\ref{G_matrices}) and (\ref{psi_AB}). It is clear that the eigenfunctions $\Psi_{n\alpha}$ are characterized by one orbital quantum number $n$, which can take four values corresponding to the states $|e_s\rangle, |e_a\rangle, |h_a\rangle, |h_s\rangle$, and the spin quantum number $\alpha$. The wave functions found in this way are, of course, orthogonal. In the next section they will be used to study the two-electron states. 

\section{Two-electron bound states}
Now consider two electrons confined in the double-well potential and find the two-particle spectrum of the bound states and their spin structure.

\subsection{Hamiltonian and calculation method}
Two electrons are described by the Hamiltonian
\begin{equation}
H(1,2)=H(1)\otimes I_4 + I_4\otimes H(2) + U(1,2)\,I_{16}\,,
\label{H12}
\end{equation}
where the operator $H$ of one argument is the one-particle Hamiltonian given by Eq.~(\ref{H}), $U(1,2)$ is the electron-electron (\textit{e-e}) interaction potential, and $I_n$ is the identity matrix.

We will diagonalize the Hamiltonian~(\ref{H12}) on the basis of the two-particle wave functions of the bound states of non-interacting electrons in the double-well potential
\begin{equation}
\Psi_{n\alpha,m\beta}^{(0)}(1,2)\!=\!\frac{1}{\sqrt{2}}\left[\Psi_{n\alpha}(1)\!\otimes\!\Psi_{m\beta}(2)\!-\!\Psi_{m\beta}(1)\!\otimes\!\Psi_{n\alpha}(2)\right],
\end{equation}
where $\Psi_{n\alpha}$ is a one-particle wave function of the double-well potential, defined by Eq.~(\ref{single_WF}) and written as a corresponding four-rank spinor; $n,m=(1,4)$ is the orbital quantum number, and $\alpha, \beta = (\uparrow, \downarrow)$ is the spin index. 

To simplify further calculations, we restrict ourselves to considering only one branch of the two-well states, namely the hole- or electronlike states. To be specific we shall consider the holelike states at a positive potential, $v>0$. The two-particle states originating from the electronlike states will be considered below in this section, but as will be seen, they are not so interesting.

With this restriction, there are only two values of the orbital quantum number which correspond to the symmetric and antisymmetric states. This simplification is justified if the difference between the energies of the antisymmetric and symmetric states is much smaller than the energy difference between the electronlike and holelike states. It is seen from Fig.~\ref{fig2} that this condition is met if the interwell distance is large enough ($d>4$ for the data shown in Fig.~\ref{fig2}).

In this case there are six two-particle basis states 
\begin{equation}
\left(\Psi^{(0)}_{s\uparrow, s\downarrow}, \Psi^{(0)}_{s\uparrow, a\downarrow}, \Psi^{(0)}_{a\uparrow, s\downarrow}, \Psi^{(0)}_{s\uparrow, a\uparrow}, \Psi^{(0)}_{s\downarrow, a\downarrow}, \Psi^{(0)}_{a\uparrow, a\downarrow}\right)^T, 
\label{2particle_basis}
\end{equation}
where $s$ and $a$ stand for the symmetric and antisymmetric states. For simplicity we will number the basis states by one index, $j=(1,6)$, in the same order as in Eq.~(\ref{2particle_basis}).

Expanding the wave function $\Psi$ of the Hamiltonian~(\ref{H12}) in this basis 
\begin{equation}
\Psi=\sum\limits_{j=1}^6 C_j\Psi^{(0)}_j\,,
\label{Psi_j}
\end{equation}
we arrive at an homogeneous equation system for $C_j$ with the following matrix:
\begin{widetext}
\begin{equation}
\begin{pmatrix}
2\varepsilon_s\!+\!U_1\!-\!\mathcal{E} & 0 & 0 & 0 & 0 & U_{16}\\
0 & \varepsilon_s\!+\!\varepsilon_a\!+\!U_2\!-\!\mathcal{E} & -U_{23} & 0 & 0 & 0\\
0 & -U_{23}^* & \varepsilon_s\!+\!\varepsilon_a\!+\!U_2\!-\!\mathcal{E} & 0 & 0 & 0\\
0 & 0 &  0 & \varepsilon_s\!+\!\varepsilon_a\!+\!U_2\!-\!U_{16}\!-\!\mathcal{E} & 0 & 0 \\
0 & 0 &  0 & 0 & \varepsilon_s\!+\!\varepsilon_a\!+\!U_2\!-\!U_{16}\!-\!\mathcal{E} & 0 \\
U_{16} & 0 & 0 & 0 & 0 &  2\varepsilon_a\!+\!U_3\!-\!\mathcal{E}
\end{pmatrix}.
\label{matrix}
\end{equation}
Here $\mathcal{E}$ is the two-particle energy; $\varepsilon_{s,a}$ are the eigenenergies of the symmetric and antisymmetric one-particle states; $U_1$, $U_2$, $U_3$, $U_{16}$, and $U_{23}$ are matrix elements of the \textit{e-e} interaction potential, which are derived from a general expression of the form
\begin{align}
U_{ij}&=\langle i=(n,\alpha;m,\beta)|U|j=(n',\alpha';m',\beta')\rangle \notag\\
&=\langle|\Psi_{n\alpha}(1)\otimes\Psi_{m\beta}(2)|U(1,2)|\Psi_{n'\alpha'}(1)\otimes\Psi_{m'\beta'}(2)\rangle -\langle|\Psi_{n\alpha}(1)\otimes\Psi_{m\beta}(2)|U(1,2)|\Psi_{m'\beta'}(1)\otimes\Psi_{n'\alpha'}(2)\rangle \notag \\
&=\delta_{\alpha,\alpha'}\delta_{\beta,\beta'}\int d^2r_1d^2r_2\left[\psi_{n\alpha}^{\dagger}(1)\psi_{n'\alpha}(1)\right]U(1,2)\left[\psi_{m\beta}^{\dagger}(2)\psi_{m'\beta}(2)\right]\notag\\
&\quad -\delta_{\alpha,\beta'}\delta_{\beta,\alpha'}\int d^2r_1d^2r_2\left[\psi_{n\alpha}^{\dagger}(1)\psi_{m'\alpha}(1)\right]U(1,2)\left[\psi_{m\beta}^{\dagger}(2)\psi_{n'\beta}(2)\right],
\label{Uij}
\end{align}
where $n,n',m,m'$ designate symmetric and antisymmetric states. In Eq.~(\ref{Uij}), the first term on the right-hand side describes the direct Coulomb interaction, while the second one corresponds to the exchange interaction. Thus we have a set of matrix elements describing the direct and exchange interactions. The specific expressions for $U_{ij}$ are greatly simplified because of the symmetry properties of the one-particle basis spinors, which are imposed by the time reversal symmetry, the $S_z$ symmetry, and the symmetry of the spinor components with respect to the spatial inversion. Finally, it turns out that the matrix~(\ref{Uij}) contains five matrix elements that determine the \textit{e-e} interaction effect on the spectrum and spin of the two-electron states: $U_1$, $U_2$ and $U_3$ describe the Coulomb repulsion of electrons in the basis states; $U_{16}$ describes both the exchange interaction in states $\Psi_{s\uparrow,a\uparrow}$ and $\Psi_{s\downarrow,a\downarrow}$, and the mixing of the states $\Psi_{s\uparrow,s\downarrow}$ and $\Psi_{a\uparrow,a\downarrow}$, and $U_{23}$ describes the mixing of states $\Psi_{s\uparrow,a\downarrow}$ and $\Psi_{a\uparrow,s\downarrow}$.

An important point is that the matrix elements are determined by nonzero components of all one-particle spinors, which can be represented in the form
\begin{equation}
\Psi_{s\uparrow}=
\begin{pmatrix}
\psi_1\\ \psi_2\\0 \\0
\end{pmatrix}\!, 
\Psi_{s\downarrow}=
\begin{pmatrix}
0 \\0\\ \psi_1^*\\ \psi_2^*\\
\end{pmatrix}\!, 
\Psi_{a\uparrow}=
\begin{pmatrix}
\phi_1\\ \phi_2\\0 \\0
\end{pmatrix}\!, 
\Psi_{a\uparrow}=
\begin{pmatrix}
0\\ 0\\ \phi_1^*\\ \phi_2^*\\
\end{pmatrix}\!,
\end{equation}
where we have taken into account a relation between spin-up and spin-down states imposed by the time reversal symmetry. The components $\psi_1$, $\psi_2$, $\phi_1$, and $\phi_2$ are calculated as described in Sec.~\ref{Sec2}. 

The matrix elements $U_{ij}$ in Eq.~(\ref{matrix}) are expressed in terms of the one-particle spinor components as follows:
\begin{align}
U_1=&\int d^2r_1d^2r_2\left[|\psi_1(1)|^2+|\psi_2(1)|^2\right]U(|r_1-r_2|)\left[|\psi_1(2)|^2+|\psi_2(2)|^2\right],\\
U_2=&\int d^2r_1d^2r_2\left[|\psi_1(1)|^2+|\psi_2(1)|^2\right]U(|r_1-r_2|)\left[|\phi_1(2)|^2+|\phi_2(2)|^2\right],\\
U_3=&\int d^2r_1d^2r_2\left[|\phi_1(1)|^2+|\phi_2(1)|^2\right]U(|r_1-r_2|)\left[|\phi_1(2)|^2+|\phi_2(2)|^2\right],\\
U_{16}=&\int d^2r_1d^2r_2\left[\psi^*_1(1)\phi_1(1)+\psi^*_2(1)\phi_2(1)\right]U(|r_1-r_2|)\left[\psi_1(2)\phi^*_1(2)+\psi_2(2)\phi^*_2(2)\right],\\
U_{23}=&\int d^2r_1d^2r_2\left[\psi^*_1(1)\phi_1(1)+\psi^*_2(1)\phi_2(1)\right]U(|r_1-r_2|)\left[\psi^*_1(2)\phi_1(2)+\psi^*_2(2)\phi_2(2)\right].
\end{align}
\end{widetext}

The solutions of a homogeneous system of linear equations with matrix~(\ref{matrix}) determine the wave functions and spectrum of the two-electron system. 

\subsection{Two-particle spectra and S-T transition}

The two-particle states defined by the matrix~(\ref{matrix}) are divided into three groups.

It is clear that both the fourth and fifth determine a separate state, $\Psi_4$ and $\Psi_5$, respectively, that is uncoupled from the others. They are determined by the spin-polarized basis states $\Psi_4=\Psi^{(0)}_{s\uparrow, a\uparrow}$ and $\Psi_5=\Psi^{(0)}_{s\downarrow, a\downarrow}$ and have the same energy
\begin{equation}
 \mathcal{E}_T\equiv \mathcal{E}_4=\mathcal{E}_5 = \varepsilon_s + \varepsilon_a + U_2 - U_{16}\,.
 \label{E_4,5}
\end{equation}
States $\Psi_4$ and $\Psi_5$ can be conventionally named tripletlike ones. 

If we speak of the classification of two-particle states by the spin, it is worth noting that in the BHZ model the total spin operator of two electrons $\hat{S^2}$ does not commute with the Hamiltonian~(\ref{H12}), and therefore, total spin is not a well-defined quantity. Thus, the conventional triplet states are not well defined. Nevertheless, the total projection of the spin onto the $z$ axis $\hat{S}_z$ is well defined, and $S_z$ is the only spin quantum number. This is quite similar to the two-electron bound states existing without confining potential~\cite{PhysRevB.95.085417}.

The other four unpolarized states split into two independent pairs, which are determined by equations corresponding to  the first and sixth lines and second and third lines of the matrix~(\ref{matrix}). 

The basis states $\Psi^{(0)}_{s\uparrow, s\downarrow}$ and $\Psi^{(0)}_{a\uparrow, a\downarrow}$ are mixed into the states $\Psi_1$ and $\Psi_6$. Their energy is equal to
\begin{equation}
\mathcal{E}_{1,6}=\varepsilon_s+\varepsilon_a +\frac{U_1+U_3}{2}\mp \sqrt{\left(\varepsilon_s\!-\!\varepsilon_a\!+\!\frac{U_1\!-\!U_3}{2}\right)^2\!+U^2_{16}}\,,
\label{E_1,6}
\end{equation}
and the wave functions are
\begin{equation}
\Psi_{1,6}=C\left(U_{16}\Psi^{(0)}_{s\uparrow, s\downarrow} - (\varepsilon_s\!+\!U_1\!-\!\mathcal{E}_{1,6})\Psi^{(0)}_{a\uparrow, a\downarrow}\right),
\end{equation}
States $\Psi_1$ and $\Psi_6$ correspond to singlet states in the case of the usual one-band spectrum. 

The basis states $\Psi^{(0)}_{s\uparrow, a\downarrow}$ and $\Psi^{(0)}_{s\downarrow, a\uparrow}$ generate the states
\begin{equation}
\Psi_{2,3}=C\left(U_{23}\Psi^{(0)}_{s\uparrow, a\downarrow} \mp |U_{23}|\Psi^{(0)}_{s\downarrow, a\uparrow}\right)
\end{equation}
with the energy
\begin{equation}
\mathcal{E}_{2,3}=\varepsilon_s+\varepsilon_a+U_2\pm|U_{23}|\,.
\label{E_2,3}
\end{equation} 
The low-energy state $\Psi_3$ corresponds to the usual unpolarized triplet state, and the high-energy state $\Psi_2$ corresponds to the singlet.

\begin{figure}
\centerline{\includegraphics[width=1.0\linewidth]{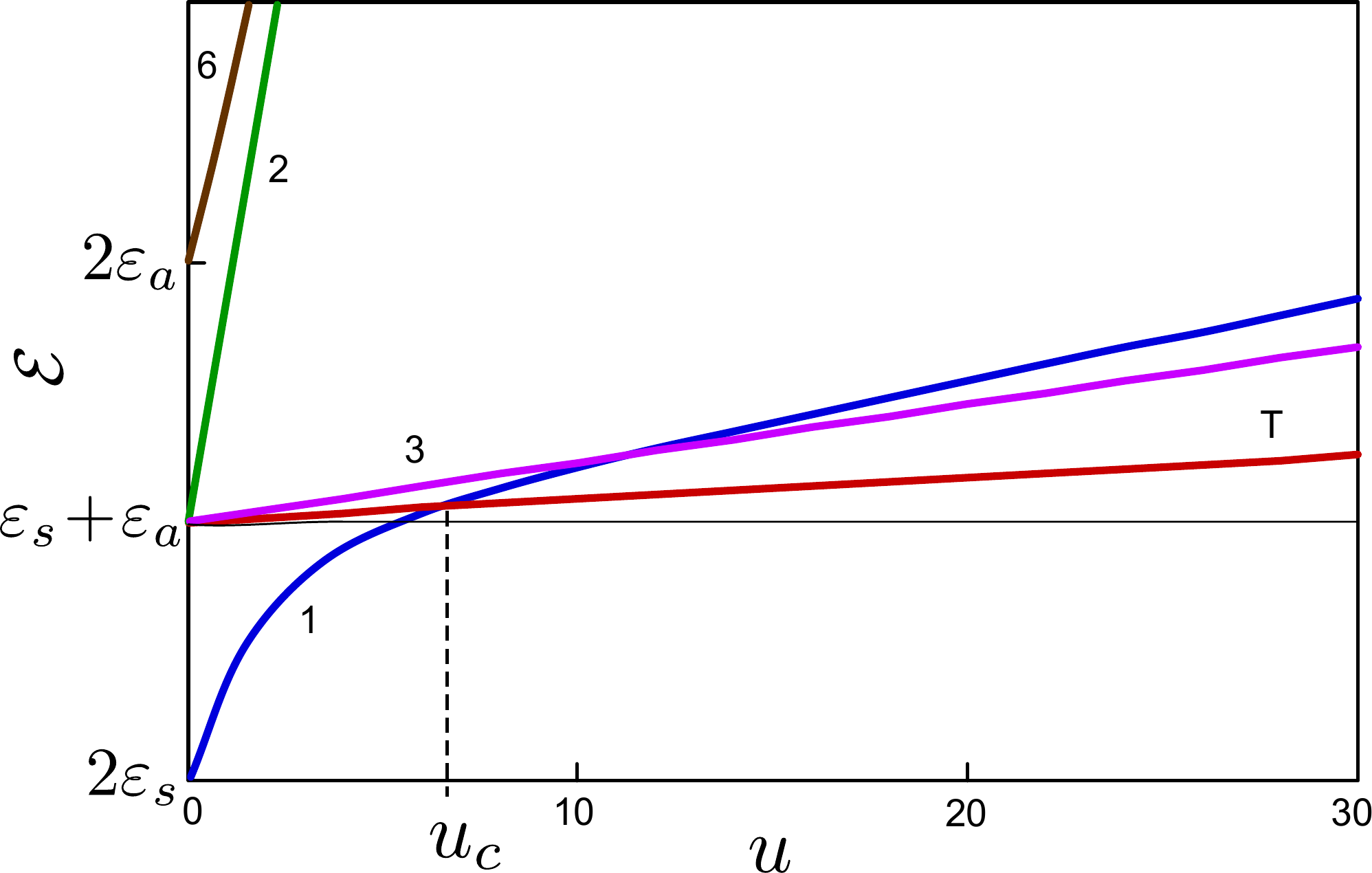}}
\caption{(Color online) Evolution of the two-electron spectrum as the \textit{e-e} interaction amplitude is changed. Lines 1, 2, 3, and 6 depict the energy of unpolarized states $\Psi_1$, $\Psi_2$, $\Psi_3$, and $\Psi_6$. Line T depicts polarized states $\Psi_4$ and $\Psi_5$. Calculations are carried out for the topological phase at $d=6$, $v=15$, $\Lambda=3$, and $a=2$. The energies of the one-particle states are $\varepsilon_s=-0.5142$ and $\varepsilon_a=-0.4786$}
\label{fig3}
\end{figure}

Thus, the total spectrum of the two-particle states is given by Eqs.~(\ref{E_1,6}), (\ref{E_2,3}), and (\ref{E_4,5}). Now it is interesting to trace the evolution of the spectrum as the \textit{e-e} interaction potential changes. Specific calculations are carried out in a simplified case of short-range \textit{e-e} interaction where the potential $U(\mathbf{r})$ is approximated by a $\delta$ function, $U(\mathbf{r})=(u/\pi)\delta(r^2)$. The results are presented in Fig.~\ref{fig3}. 

The energy of the polarized states is seen to be slowly increasing with $u$, with their spin and pseudospin structure being independent of $u$. In contrast, the energy of the unpolarized states and their pseudospin structure varies appreciably with the \textit{e-e} interaction potential.

The most interesting feature of the two-particle spectra is the crossing of the energy levels of the unpolarized singlet state $\Psi_1$ and the polarized states $\Psi_4$  and $\Psi_5$ at some value of the \textit{e-e} interaction potential $u=u_c$. As a result, the ground state, which is the singlet at small $u$, becomes a triplet when $u>u_c$. 

\begin{figure}
\centerline{\includegraphics[width=1.0\linewidth]{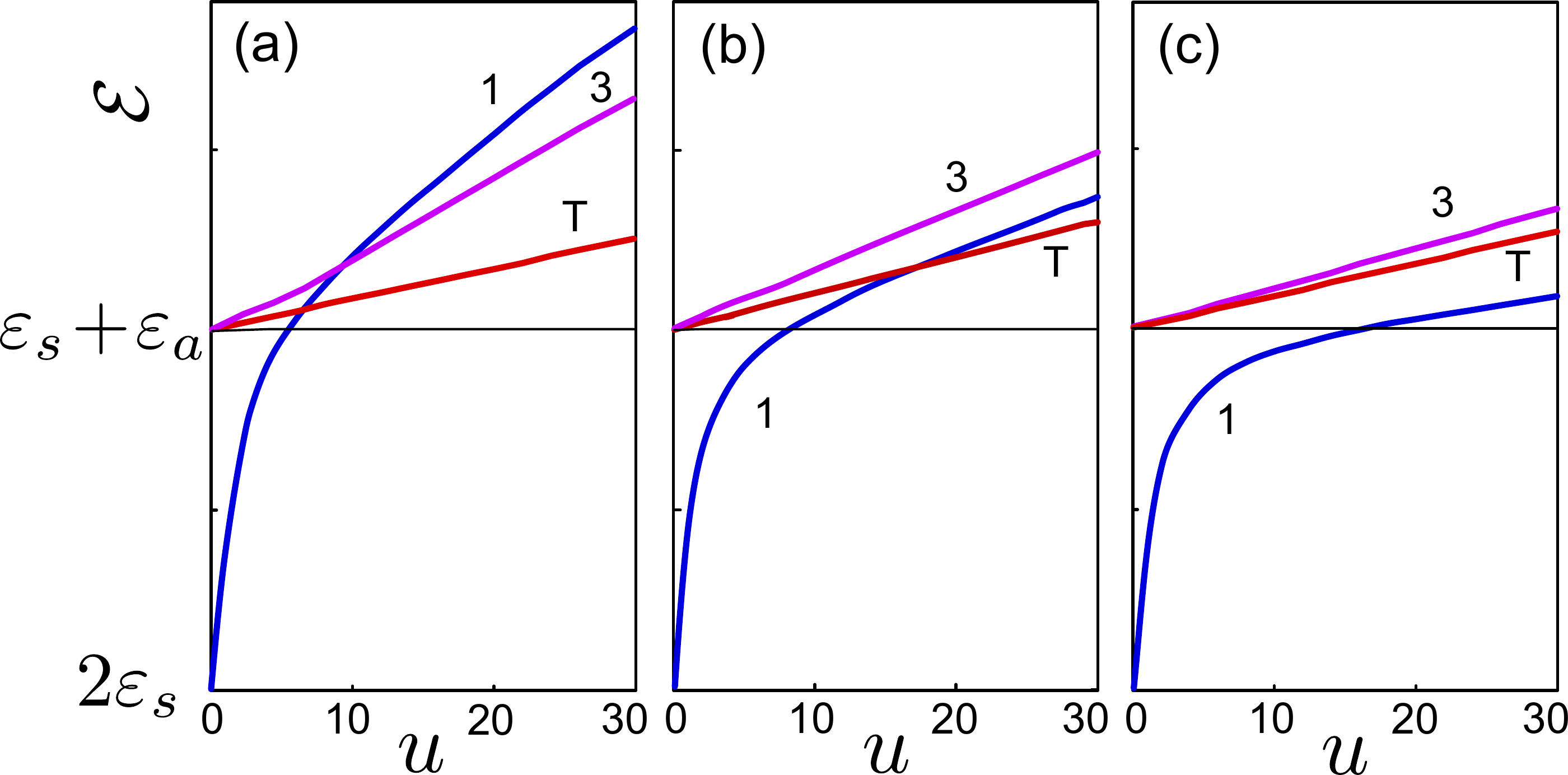}}
\caption{(Color online) Evolution of the two-electron spectrum as the \textit{e-e} interaction amplitude is changed for a variety of the well potentials. The low-lying singlet $\mathcal{E}_1$ (line 1), polarized triplets $\mathcal{E}_T$ (line T), and unpolarized tripletlike state $\mathcal{E}_3$ (line 3) are shown. The numerical data are as follows: (a) $v=15$, $\varepsilon_s=-0.5142$, and $\varepsilon_a=-0.4786$; (b) $v=18$, $\varepsilon_s=-0.2803$, and $\varepsilon_a=-0.2473$; and (c) $v=25$, $\varepsilon_s=0.0515$, and $\varepsilon_a=0.0849$. Other parameters are $d=6$, $\Lambda=3$, and $a=2$.}
\label{fig4}
\end{figure}

Shown in Fig.~\ref{fig3} the intersection of a low-lying singlet and triplet occurs at $v=15$. The intersection exists in a wide region of $v$, but with increasing $v$  the energy of the unpolarized singlet state $\mathcal{E}_1$ decreases relative to that of the polarized states $\mathcal{E}_{T}$, and at a critical value $v=v_c$ the intersection disappears, as shown in Fig.~\ref{fig4}. 

The intersection of a low-lying singlet with a triplet occurs in a rather wide range of system parameters (they are the potential of the quantum well $v$, the \textit{e-e} interaction potential $u$, and the distance between the wells $d$) when any of them is changed. The most interesting for realization is, apparently, the S-T transition under a change in the potential of quantum wells, which can be realized by means of a gate. Therefore, we have studied the conditions for the S-T transition to appear when $v$ and $u$ are changed for a fixed value of $d$.

\begin{figure}
\centerline{\includegraphics[width=0.9\linewidth]{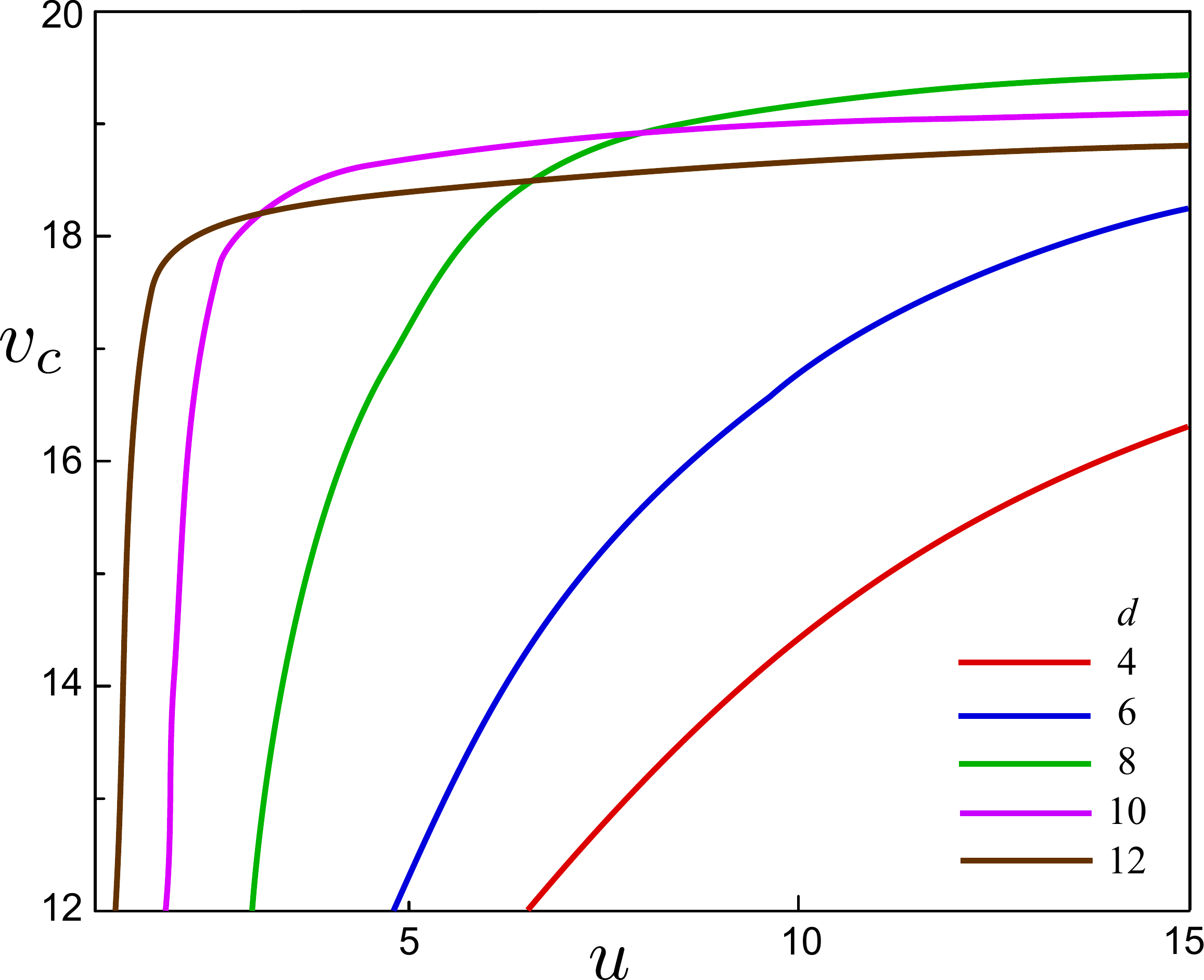}}
\caption{(Color online) Critical potential of the quantum wells $v_c$ as a function of the \textit{e-e} interaction potential $u$ for a variety of interwell distances $d$. In the region $v<v_c$ the ground state is formed by the triplet states. The parameters used in the calculations are $a=2$, $\Lambda=3$.}
\label{fig5}
\end{figure}

The critical value of the potential $v_c$ as a function of the interaction potential $u$ is presented in Fig.~\ref{fig5} for a variety of the interwell distances $d$. If we consider Fig.~\ref{fig5} as the $u$-$v$ plane, each line divides the map into two regions for a given $d$. Below the line the ground state is a tripletlike state formed by one of the polarized triplet states or their superposition. Above the line the ground state is the singlet. Thus, if we increase the well potential at a given \textit{e-e} interaction potential, the initial tripletlike state switches to the singlet state at a critical value of $v$. It is interesting that there is a critical value of $u$ below which no S-T transition is possible, with this critical $u$ being dependent on the interwell distance.

The energy level $\mathcal{E}_3$ of the unpolarized tripletlike state $\Psi_3$ can also cross the singlet level $\mathcal{E}_1$; however, the unpolarized triplet never becomes the ground state, as seen from Fig.~\ref{fig4}.

The scenario of the S-T transition can be understood at a qualitative level if we consider variations of the energy levels of both the singlet and triplet states with increasing \textit{e-e} interaction potential $u$. The energy of the spin-polarized states $\mathcal{E}_T$ grows rather slowly with $u$ because, as seen from Eq.~(\ref{E_4,5}), the \textit{e-e} repulsion energy $U_2$ is partially compensated by the exchange energy $U_{16}$. The singlet energy $\mathcal{E}_1$ rises much faster as long as $u$ is small since $\mathcal{E}_1$ grows only because of the \textit{e-e} repulsion energy $U_1$. However, with increasing $u$ it becomes important that the low-lying basis state $\Psi_{s\uparrow,s\downarrow}$ is mixed with the excited singlet state $\Psi_{a\uparrow,a\downarrow}$ due to the exchange interaction $U_{16}$. As a result, the variation of this energy term becomes more complicated. According to Eq.~(\ref{E_1,6}), in the limit of large $u$ the low-energy singlet energy is approximated as
\begin{equation}
\mathcal{E}_1\simeq \frac{U_1+U_3}{2} - \sqrt{\left(\frac{U_1\!-\!U_3}{2}\right)^2\!+U^2_{16}}\,.
\end{equation}

It is seen, that at large $u$ both the further growth of the singlet energy and its fall with $u$ are possible depending on specific properties of the exchange interaction matrix element $U_{16}$. In particular, in the limit $u\to\infty$, the critical condition for the singlet energy to grow is
\begin{equation}
U_1 U_3 > U^2_{16}\,.
\label{criterionN}
\end{equation}

Of course, this is a necessary condition for the crossing to appear. The crossing appears when
\begin{equation}
U_2-U_{16}\le \frac{U_1+U_3}{2}-\sqrt{\left(\frac{U_1-U_3}{2}\right)^2+U^2_{16}}\,.
\label{criterionS}
\end{equation}

There are no \textit{a priori} prohibitions that prevent the fulfillment of these conditions. The question can be posed as follows: is there no prohibition on the form of the wave function for which the matrix elements of $U_{ij}$ can satisfy these conditions? We can state with certainty only that the wave function must have pseudospin components. This conclusion was obtained from the consideration of a limiting case of weakly coupled QDs when the symmetric and antisymmetric one-particle wave functions $\Psi_{s\uparrow}$, $\Psi_{s\downarrow}$, $\Psi_{a\uparrow}$, and $\Psi_{a\downarrow}$ can be represented as $\Psi_{s,a}(\mathbf{r})=2^{-1/2}\left[\Psi^{(1)}(\mathbf{r}-\mathbf{R}_A)\pm \Psi^{(1)}(\mathbf{r}-\mathbf{R}_B) \right]$, where $\Psi^{(1)}(\mathbf{r})$ is the wave function in the single quantum well. In this way we have found that these conditions are never satisfied if only one pseudospin component is nonzero. 

At this point we should stress the important difference in the evolution of the low-lying singlet level $\mathcal{E}_1$ with $u$ in the cases of the two-band model we are considering and the usual one-band one where the pseudospin is absent. In the one-band model the mixing of the low-lying basis singlet $\Psi_{s\uparrow, s\downarrow}$ and the excited singlet $\Psi_{a\uparrow, a\downarrow}$ results asymptotically in the saturation of the $\mathcal{E}_1$ growth because the exchange energy exactly compensates the repulsion energy. This happens at all system parameters ($v$ and $d$).

In the two-band model the situation essentially changes so that $\mathcal{E}_1$ can either increase or decrease with $u$ depending on the system parameters which determine the wave functions and the matrix element $U_{16}$. When $U^2_{16}>U_1 U_3$ (in terms of the one-band model this could be interpreted in such a way that the exchange energy is greater than the repulsion energy) the low-lying singlet energy $\mathcal{E}_1$ asymptotically decreases with $u$, and the ground state is the singlet. When the exchange energy is small, $U^2_{16}<U_1 U_3$, the low-lying singlet energy $\mathcal{E}_1$ grows with $u$. If in addition the condition~(\ref{criterionS}) is met, the singlet crosses the triplet level, and the S-T transition can occur.

To conclude the consideration of the S-T transition in the topological phase we estimate whether the approximation used in the calculations is justified. Using the basis of the noninteracting two-electron states to diagonalize the Hamiltonian implies that we act within the degenerate perturbation theory~\cite{bir1974symmetry}. This imposes a restriction on the value of the \textit{e-e} interaction potential. Thus, we have to check, whether the interaction potential in the crossing point of the energy levels of the polarized and unpolarized states is so small that the perturbation theory is applicable. In order to clarify this, we have calculated the energy increments due to the \textit{e-e} interaction for all the terms at the S-T transition point $\Delta\varepsilon_j=\varepsilon_j(u)- \varepsilon_j(u=0)$ and compared the maximal value $\mathrm{max}[\Delta\varepsilon_j]$ with the energy difference between the holelike and electronlike states of noninteracting electrons $\Delta\varepsilon_{he}$. The calculations give the following result for the set of parameters $a=2$, $\Lambda=3$, $v=15$, and $d=8$. The singlet and triplet terms intersect at $u=4$. In the crossing point, the ratio $\mathrm{max}[\Delta\varepsilon_j]/\Delta\varepsilon_{he}$ is lower than 0.2, which is enough for the perturbation theory to give a qualitatively correct result. 

Further numerical calculations of the ratio $\mathrm{max}[\Delta\varepsilon_j]/\Delta\varepsilon_{he}$ as a function of the distance between the wells have shown that this ratio decreases with increasing $d$, which evidences that the \textit{e-e} interaction effectively decreases with $d$. Thus, the perturbative consideration of the \textit{e-e} interaction is justified at large enough $d$.

We have studied also the two-electron bound states constructed on the basis of the single-particle bound states of the electron type for the positive potential $v$. Their spectra are shown in Figs.~\ref{fig1} and~\ref{fig2}. However, it turns out that no crossing of the tripletlike and singletlike states occurs in this case.

\begin{figure}
\centerline{\includegraphics[width=1.0\linewidth]{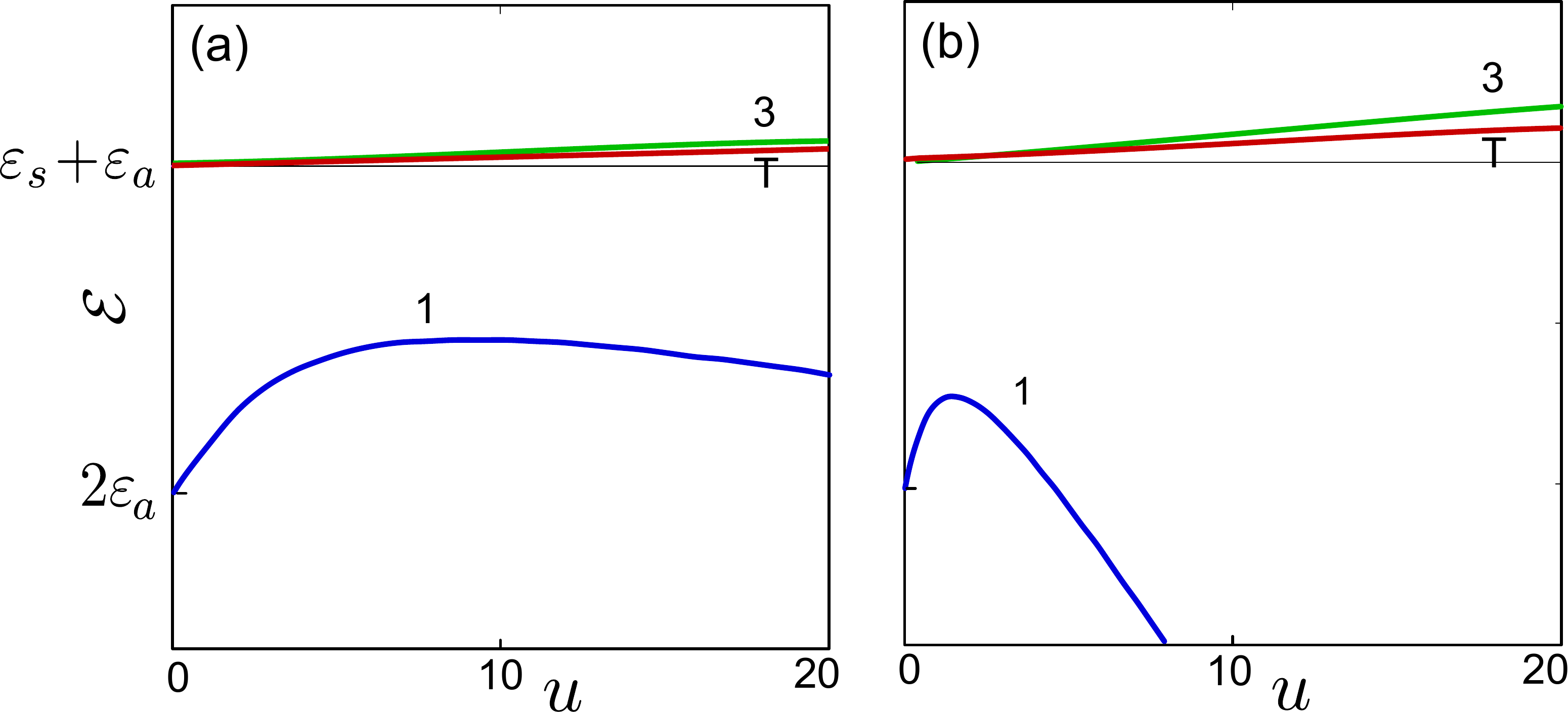}}
\caption{(Color online) Evolution of the two-electron spectrum as the \textit{e-e} interaction amplitude is changed for two values of the well potential in the topologically trivial case. The low-lying singlet $\mathcal{E}_1$ (line 1), polarized triplets $\mathcal{E}_T$ (line T), and unpolarized tripletlike state $\mathcal{E}_3$ (line 3) are shown. The numerical data are as follows: (a) $v=8$, $\varepsilon_s=-0.698$, and $\varepsilon_a=-0.83$; and (b) $v=12$, $\varepsilon_s=0.248$, and $\varepsilon_a=0.182$. Other parameters are $d=6$, $\Lambda=3$, and $a=2$.} 
\label{fig6}
\end{figure}

In order to clarify how the topology of the band spectrum of the host material manifests itself in the spectrum and spin of the two-electron states, we have studied the situation of the topologically trivial material using the same approaches. In this case, the one-particle spectrum contains only one bound state (holelike state for $v>0$ and electronlike one for $v<0$). In addition, it is essential that the arrangement of the energy levels of the symmetric and antisymmetric states is the opposite of the corresponding holelike or electronlike states in the topologically nontrivial case. This means that the low-lying basis state is antisymmetric, and the high-lying one is symmetric.

The calculations have shown that in the trivial case there are no crossings of the two-particle terms and no S-T transition occurs. The energy spectrum of the low-energy singlet state and the triplet states is illustrated in Fig.~\ref{fig6} for two values of the potential $v$. Analysis of the conditions~(\ref{criterionN}) and (\ref{criterionS}) shows that even the necessary condition~(\ref{criterionN}) does not hold at any parameters of the system.

\section{Concluding remarks}
The exchange interaction is known to play a key role in the formation of the spin structure of two-particle states in double-QD structures, as well as in processes of spin manipulation with using gate voltages that control the potential relief. A change in the simple single-band spectrum caused by adding new orbital degrees of freedom leads to an essential change in the exchange interaction and, correspondingly, in the spin structure of the two-particle states. This fact was demonstrated, for example, in a quasi-two-dimensional system, where adding dimension-quantization levels led to lowering of the triplet state relative to its counterpart singlet state~\cite{PhysRevA.81.022501}, or in a double-well structure, where the effect of excited orbitals was found~\cite{PhysRevB.97.045306}. In the present paper we have shown that more dramatic effects arise due to the orbital degrees of freedom described by pseudospin in host materials with a topologically nontrivial band spectrum. 

We have studied two-electron states confined in a double-well potential in the 2D topological insulator, described by the BHZ model, and found that there is a wide range of system parameters in which the ground state is formed by polarized triplet states. Outside this range the ground state is a singlet. We have established the critical conditions under which the ground state can be transformed from singlet to triplet and vice versa by changing the potential $v$ of the wells. The S-T transition occurs at a critical value of $v=v_c$, which depends on the \textit{e-e} interaction potential amplitude $u$ so that $v_c$ increases with $u$. 

The S-T transition occurs because of the peculiarities of the \textit{e-e} interaction effect on the low-lying singlet and triplet energies in a material with an inverted two-band spectrum. The triplet energy weakly increases with the \textit{e-e} interaction potential since the repulsion energy is greatly compensated by the exchange interaction. On the other hand, the energy of the singlet, under certain conditions when the S-T transition occurs, significantly increases with the \textit{e-e} interaction potential since the exchange energy slightly compensates the repulsion energy. As a result, the crossing of the singlet and triplet levels becomes possible.

Our consideration is restricted by a simplified model system of two identical quantum wells with a short-range potential for the topological phase with the hybridization parameter of the electron and hole bands $a\ge 2$. The \textit{e-e} interaction is approximated by a $\delta$ function. More cumbersome and time-consuming calculations have shown that the main result (the crossing of the singlet and triplet terms in some range of the system parameters) persists as we deviate from this model. 

So a generalization to the \textit{e-e} interaction potential of finite radius shows that the crossing of the singlet and triplet levels persists in any case as the interaction radius is small in comparison with the distance between the wells. 

An asymmetry of double-well potential more strongly affects the behavior of the singlet and triplet terms. A small potential difference $\delta v$ between the quantum wells leads to an avoided crossing of the singlet level $\mathcal{E}_1$ and the level $\mathcal{E}_3$ of the unpolarized tripletlike state $\Psi_3$ in contrast to the crossing shown in Fig.~\ref{fig3}. Further increasing $\delta v$ leads to the disappearance of the crossing of the singlet and polarized triplet terms which happens when $\delta v$ is comparable with the energy difference between the bound state levels.

At the critical point of the S-T transition, the energy levels of the polarized and unpolarized states intersect. In this work we did not take into account any spin-dependent interactions. We expect that inclusion of a spin-orbit interaction due to breaking down the inversion symmetry to result in an avoided crossing of these terms, although we recognize that the question of the spin structure of two-electron states in the presence of the spin-orbit interaction is a challenging problem that deserves a separate study.
 
\acknowledgments
This work was supported by the Russian Science Foundation under Grant No.~16-12-10335.

\bibliography{paper}

\end{document}